\documentclass[10pt,conference]{IEEEtran}
\IEEEoverridecommandlockouts

\usepackage{graphicx}
\usepackage{textcomp}
\usepackage{xcolor}
\usepackage{lscape}
\usepackage{float}
\usepackage{multirow}
\usepackage{tcolorbox}
\usepackage{booktabs}
\usepackage{caption}
\usepackage{subcaption}
\usepackage{balance}
\usepackage{algorithm}
\usepackage{algorithmic}
\usepackage{soul}
\usepackage{hyperref}

\def\BibTeX{{\rm B\kern-.05em{\sc i\kern-.025em b}\kern-.08em
		T\kern-.1667em\lower.7ex\hbox{E}\kern-.125emX}}

\begin{document}

\title{Microusity: A testing tool for Backends for Frontends (BFF) Microservice Systems}


\author{
\IEEEauthorblockN{Pattarakrit Rattanukul\textsuperscript{*}, Chansida Makaranond\textsuperscript{*}, Pumipat Watanakulcharus\textsuperscript{*}, Chaiyong Ragkhitwetsagul\textsuperscript{*},\\ Tanapol Nearunchorn\textsuperscript{$\dagger$}, Vasaka Visoottiviseth\textsuperscript{*}, Morakot Choetkiertikul\textsuperscript{*}, Thanwadee Sunetnanta\textsuperscript{*}}
\IEEEauthorblockA{\textsuperscript{*}\textit{Faculty of ICT, Mahidol University}, Thailand, \textsuperscript{$\dagger$}\textit{Lineman Wongnai}, Thailand}
}

\maketitle

\begin{abstract}
The microservice software architecture is more scalable and efficient than its monolithic predecessor. 
Despite its increasing adoption, microservices might expose security concerns and issues that are distinct from those associated with monolithic designs. 
We propose Microusity, a tool that performs RESTful API testing on a specific type of microservice pattern called back end for front end (BFF). 
We design a novel approach to trace BFF requests using the port mapping between requests to BFF and the sub-requests sent to back-end microservices. 
Furthermore, our tool can pinpoint which of the back end service causing the internal server error, which may lead to unhandled errors or vulnerabilities.
Microusity provides an error report and a graph visualization that reveal the source of the error and supports developers in comprehension and debugging of the errors. 
The evaluation of eight software practitioners shows that Microusity and its security test reports are useful for investigating and understanding problems in BFF systems. 
The prototype tool and the video demo of the tool can be found at \url{https://github.com/MUICT-SERU/MICROUSITY}.
\end{abstract}

\begin{IEEEkeywords}
microservices, API security, testing, fuzzing
\end{IEEEkeywords}

\section{Introduction} 
Microservice architecture has been increasingly adopted \cite{ibmmarketdevelopmentandinsights} and is frequently used when building modern software. One of the benefits of the microservice architecture is its scalability and modularity~\cite{ibm2021}. The developers can adopt a microservice pattern that is suitable to their business, such as an aggregator pattern, chained pattern, proxy pattern~\cite{Raj2017}, micro front-end pattern, and the backends for frontends (BFF) pattern~\cite{Newman2021}. The BFF pattern provides an API endpoint as a middleman for the client, i.e., the front end, to fetch the data from itself rather than fetching directly from the back-end microservices. Multiple BFFs can be created to support several target devices of the same system to support different types of data used by such devices, such as a BFF for mobile applications and a BFF for the desktop website~\cite{Newman2021}. Thus, using the BFF pattern is useful for the service with different client platforms. However, when the error occurs in the back-end microservices, it is hard to trace which service causes the error when the data is passed via the BFF. Furthermore, without proper implementation, back end problems (i.e., stack trace, error exception) may be transmitted back to the client, exposing sensitive data to the attackers.

Existing tools and techniques~\cite{Chondamrongkul2020,Taya2022,Viglianisi2020,Karlsson2020,Arcuri2021} that perform RESTful API testing can only test at the API endpoints, but cannot trace the execution after the endpoint. In the case of the BFF pattern, when an endpoint (i.e., BFF) returns an error, the developers need to manually perform the checking to identify which back-end microservice(s) is causing the error.

In this paper, we introduce \textbf{Microusity}, a tool that performs RESTful API testing of BFF microservices. The tool traces requests processed by BFF, creates main-request to sub-request mapping, and provides test reports, in both textual and visualization formats, to help developers to trace and fix issues. This paper makes the following key contributions.
\begin{enumerate}
    \item BFF API fuzzing and request tracking: A novel approach to map the request coming into the BFF to the requests sent to the back-end microservices. 
    \item A graph-based visualization to help the developers in comprehension and in debugging the errors caused by the back-end microservices.
\end{enumerate}

\section{Background}\label{sec:RelatedWork}
In this section, we discuss the background knowledge and the existing tools that are related to Microusity. 

\textbf{RESTful APIs.}
A Representational State Transfer (REST) application programming interface (API) is a web service that offers functionalities via HTTP. It has been adopted widely by industry~\cite{Arcuri2021}. Microservices usually adopt RESTful APIs as a communication method between their services.

\textbf{Microservices and their security.}
Microservice is a software architecture that breaks down an application into several decoupled components that may be deployed independently of one another~\cite{Newman2021}. The services communicate with one another via multiple methods such as RESTful APIs, event streaming, or message brokers~\cite{ibm2021}.
Furthermore, each microservice intends to be an autonomous development and run-time decision-making unit. Therefore, microservice have seen widespread use in practice~\cite{Dragoni2017,Pahl2016,Thones2015,Jamshidi2018}. Several software and service organizations, including Netflix~\cite{netflix}, Soundcloud~\cite{SoundCloud,SoundCloudThoughtworks}, and Uber~\cite{gluck_2020}, have embraced microservice architecture as a substitute for their older monolithic approaches. It is expected to grow further with the 5-year forcasted growth rate during 2022--2027 of 15.7\%~\cite{ltd_2022}.

Nevertheless, threats to the security of microservice architectures are becoming more prevalent. According to the study from Hannousse and Yahiouche~\cite{Hannousse2021}, microservices suffer from security breaches by user-based, data, infrastructure, and software attacks. Esposito et al.~\cite{Esposito2016} report that microservices can introduce more attack surfaces due to the larger number of independent services compared to their monolithic counterpart. 
As a result, the lack of a testing support framework in the RESTful API of the microservice system might cause serious issues since the RESTful API plays a huge role in integrating microservices together ~\cite{Yarygina2018}. 

\textbf{Backends for Frontends (BFF).}
BFF is one of the patterns for microservices that is designed to provide several gateways for the front ends rather than having only one API gateway for every front end. It allows fine-tuning to a specific front-end user interface (UI) such as mobile and web applications since they may have different requirements~\cite{Newman2021}. 
BFF works as an interface between the front end and the back-end microservices. It receives a request from a front end and dispatches the request or creates requests to several back-end microservices to retrieve data and aggregate the result back to the front end~\cite{Harms2017}. In this paper, we call the request to the BFF as the \textit{main request} and those requests that propagate from the BFF to back-end services as \textit{sub-requests}. This usually creates a one-to-many mapping between the main request, created by the client, to the sub-requests created by the BFF. By not carefully checking for responses that are sent from the back-end microservices, the BFF can leak sensitive information such as programming exception messages. This information can be beneficial to attackers and lead to API attacks targeting a module that caused the exception. 
Moreover, identifying the back-end service that causes an error could be troublesome in BFF. When an API request is sent to the BFF, this request is passed to several back-end services. If one of the back-end services fails and returns the 500 HTTP status code, we cannot know which of the services fail since we can only get the unsuccessful result from the BFF response.

\textbf{RESTler} is a stateful RESTful API fuzzing tool~\cite{atlidakis_godefroid_polishchuk_2019}. RESTler automatically tests RESTful APIs for finding bugs related to security and reliability issues based on the responses from APIs. It generates test requests by compiling Swagger or OpenAPI specifications and inferring the producer and consumer dependencies. In addition, RESTler can identify the state of the test sequence and use the response from previous test results to generate new test requests to find more bugs. 

\textbf{Zeek} (formerly known as Bro) is a network monitoring tool specialized in event logging and powerful event handling. It is extensible with its own programming language \cite{Zeek}. Zeek can be used to perform network traffic analysis and create alerts to the user by using scripts.




\begin{figure*}[tb]
    \centering
    \includegraphics[width=1.9\columnwidth]{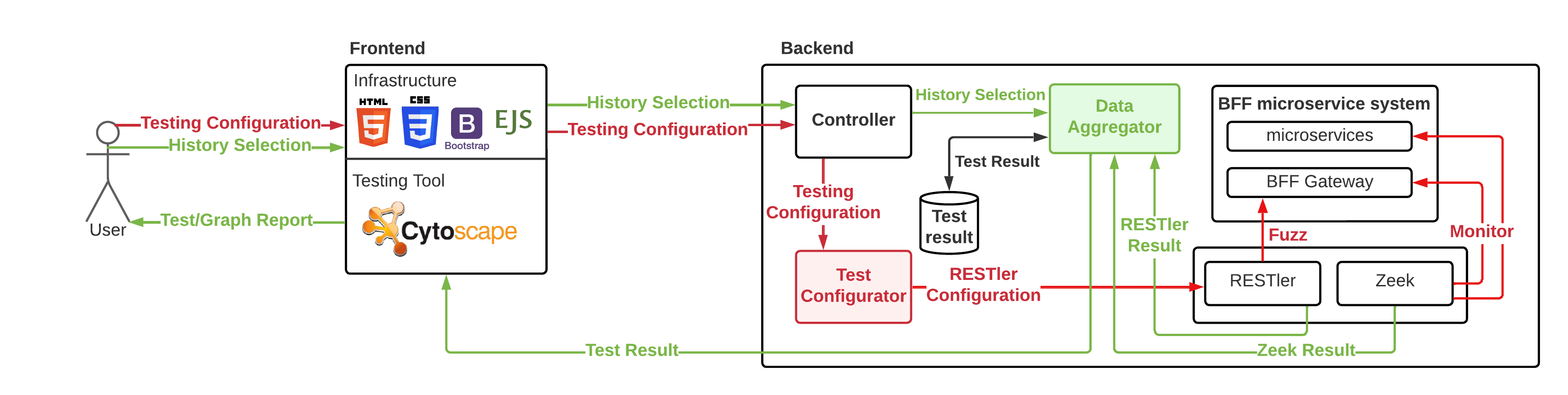}
	\caption{System architecture and workflow of Microusity}
	\label{fig:system_architecture} 
\end{figure*}

\section{The Microusity Tool}\label{sec:microusity}
Microusity is an automated API testing tool that targets BFF microservice systems. In this section, we present the Microusity system design, the approach of API fuzzing and request tracking, and the reporting mechanism. 


\subsection{System Design}
The system architecture of Microusity is depicted in Figure \ref{fig:system_architecture}. Microusity's back end is composed of three components. The first component is a \textit{Controller}. The controller reads the test configuration and handles the data flow between the front end and back end of the tool allowing the user to query test history. The second component is the \textit{Test Configurator}. Test Configurator incorporates the custom configuration that the user created. This configuration is used to adjust the test coverage of the target BFF system and also to modify the test inputs. The third component is the \textit{Data Aggregator}.  Data Aggregator processes the data collected from the test execution
and generates the test result mapping between main requests to BFF and their sub-requests. This mapping result is stored in data storage and sent to the tool's front end for processing as a test report and graph report. 
For Microusity's front end, HTML, CSS, Bootstrap, and EJS are used. The graph report from the API testing is visualized by Cytoscape.js. 

Two main components of Microusity that handle the testing part include RESTler and Zeek. We use RESTler as the RESTful API fuzzing engine and we use Zeek to intercept the network log between the BFF and their back-end services. Microusity monitors HTTP requests and responses that occur by RESTler's fuzzing in order to trace the execution of each request and to locate responses that contain errors, i.e., responses that contain HTTP response status code of client error responses (400–499) or server error responses (500–599).

\subsection{BFF API Fuzzing and Request Tracking}


We propose a novel approach for BFF API fuzzing and request tracking in this paper. Figure \ref{fig:system_architecture} is used to explain the approach. 
First, the testing configuration is supplied by the user through the web interface and passed to the Controller. The Controller passes the testing configuration to the Test Configurator, which identifies the target system to test. Next, RESTler will be executed with the provided configuration to fuzz the APIs of the tested system while Zeek monitors the communications between the BFF and the back-end microservices. After fuzzing, the Data Aggregator collects the test results from RESTler and the network monitoring log from Zeek. The mapping is performed and the final API test results are kept in the Test Result database. Lastly, the results are sent back to the front end to generate security and graph reports.

Microusity leverages the fuzzing capability from RESTler which generates the requests from the targeted BFF's OpenAPI specification. Once a request is validated, RESTler will attempt to modify the request into a malformed one by taking the BFF's state into account.
After testing is completed, Microusity reads the log and creates the main request and sub-request mapping based on the network port in the requests as described in Algorithm~\ref{alg:mapping}. For each test input generated by RESTler, Zeek collects all the requests created in the system. Then, based on the order of requests in the Zeek log (sorted chronologically), each request is checked whether it is the original request coming to BFF (i.e., by having the destination to the BFF's host and port) or the sub-request generated by BFF. Then, the sub-requests are mapped to the original request. The process keeps repeating until all the requests in the Zeek log are processed. 

\begin{algorithm}[tb]
\caption{Mapping of main-request and its sub-requests}\label{alg:mapping}
\footnotesize
\begin{algorithmic}
\REQUIRE $\mathrm{zeeklog}$ is sorted chronologically.
    \FOR {$ \mathrm{request} \in \mathrm{zeeklog}$}
        \IF {$\mathrm{request}.port = \mathrm{BFF\_PORT}$}
            \STATE $\mathrm{request\_map}.insert(\mathrm{request})$
            \STATE $\mathrm{BFF\_REQUEST} \gets \mathrm{request}$
        \ELSE
            \STATE $\mathrm{request\_map}.get(\mathrm{BFF\_REQUEST}).\mathrm{add\_subrequest}(\mathrm{request})$
        \ENDIF
    \ENDFOR
\end{algorithmic}
\end{algorithm}


\subsection{Reporting Mechanism}
Microusity offers two types of reports: the error report and the graph report. We explain them below.

\subsubsection{Error Report}

\begin{figure}[tb]
	\centering
	\includegraphics[width=0.95\columnwidth]{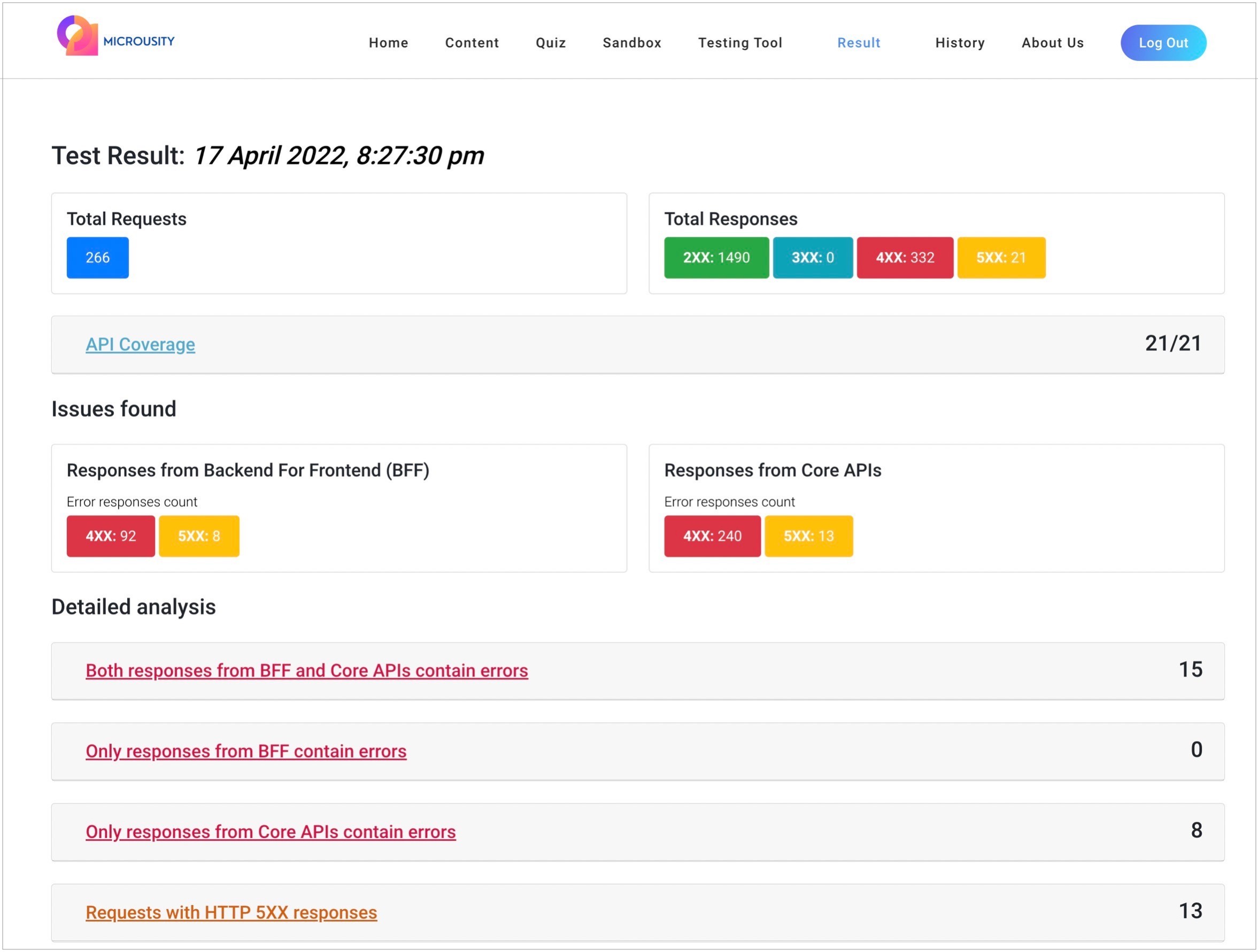}
	\caption{Example of the error report}
	\label{fig:testReportExample} 
\end{figure}

The error report is divided into three main sections as shown in Figure~\ref{fig:testReportExample}. The first section is the overall test summary. This section tells
the API coverage, the number of total requests sent to the BFF, the total of responses, and
HTTP status code from all responses, including sub-responses. The second section shows
the total number of error responses from the BFF and back-end microservices. This can help the user
understand how many errors are occurred from the BFF or back-end microservices. The last section is the request sequence that contains issues found by Microusity. This section is grouped into four
different categories. The first category contains the request sequences in which their response contains an exception leakage in both of main response from the BFF and sub-responses from the back-end microservices. The second category contains the request sequence that only the main response from the BFF contains the exception leakage. The third category contains the sequence that only the sub-responses from back-end microservices contain exception leakage. The last category includes sequences that contain the HTTP 5xx (i.e., HTTP 500--599 indicating a server error) as their return status code.
We choose to monitor HTTP 5xx responses because they indicate that an unhandled issue occurs in the back-end microservices, and potentially contains a security vulnerability\footnote{Internal Server Error (HTTP 500) is a response from a server when it finds an error that it does not know how to handle. More info: \url{https://developer.mozilla.org/en-US/docs/Web/HTTP/Status}}.

\subsubsection{Graph Report}

\begin{figure}[tb]
	\centering
	\includegraphics[width=\columnwidth]{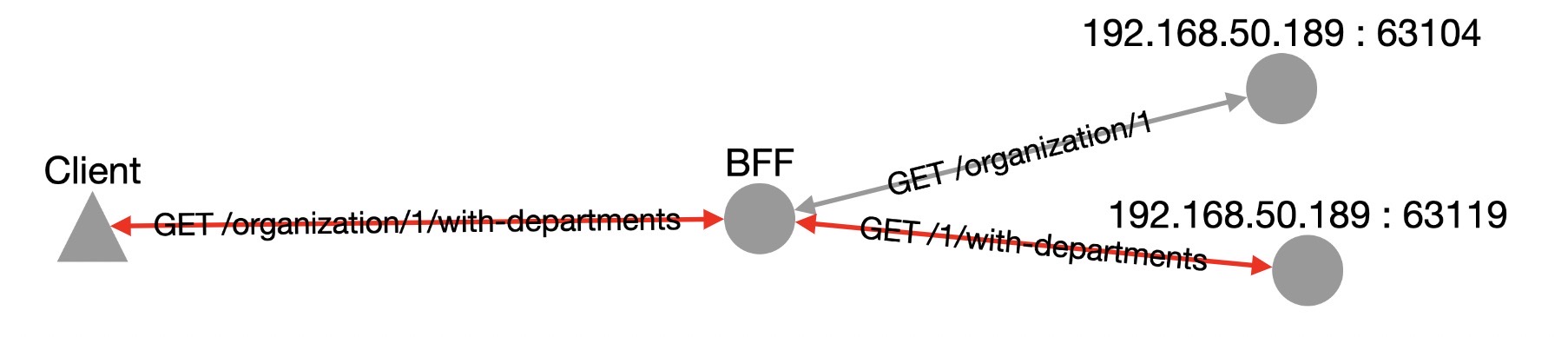}
	\caption{Example of the graph report}
	\label{fig:graphReportExample} 
\end{figure}

Microusity offers a graph report to visualize the relationship between the main request and sub-requests and their responses. Figure \ref{fig:graphReportExample} shows an example of the graph report. It depicts an API request and responses that pass through the BFF. The triangle node represents the client. The arrow on the left is the main request and main response from Microusity to the BFF, while the arrows on the right are the sub-requests that the BFF creates after receiving the client request. Each request is labeled by IP address and port number. Microusity provides this graph visualization for the sequence that contains the issues shown in the error report to give additional information and easy-to-trace connections between the main request and the sub-requests to the developers. The requests which contain exception message leakage or HTTP 5xx are highlighted in red and the user can expand the request arrow to show more information such as the request's or response's bodies and header.

\section{Evaluation}
We performed a user evaluation to evaluate the ease of understanding and the usefulness of Microusity using a demonstration and an interview. 
We recruited eight full-time software engineers from four different companies. 
The demographic of the interviewed participants is shown in Table~\ref{tab:live_interview_ppl}. The participants are software architects and software engineers with working experience from 1 to 7 years. They all have worked with microservice systems ranging from half a year to 6 years. We explained the Microusity tool and its API testing concept to the participants. Then, for three participants, we demonstrated the tool execution on a simple BFF project\footnote{https://github.com/piomin/sample-spring-microservices-new} with the security test report and graph report. For the other five participants, we demonstrated the tool execution on their company's BFF system. Then, we asked them 7 questions to rate the usability and usefulness of the system using the 5-level Likert scale. The full list of interview questions is on our study website\footnote{\url{https://github.com/MUICT-SERU/MICROUSITY}}.

\begin{table}[tb]
    \centering
    \footnotesize
    \resizebox{\columnwidth}{!}{%
    \begin{tabular}{lrr}
    \toprule
    Position & Working Exp. & Exp. with microservices \\
    \midrule
    Solution Architect & 7 years & 6 years  \\
    Senior Software Engineer & 5 years    & 3 years   \\
    Software Engineer        & 2.5  years & 2.5 years \\
    Software Engineer        & 3 years    & 2 years   \\
    Software Engineer        & 3 years    & 1.5 years \\
    Software Engineer        & 1 years    & 1 years   \\
    Senior Software Engineer & 7 years    & 0.7 year  \\
    Software Architect       & 3 years    & 0.5 year \\
    \bottomrule
    \end{tabular}
    }
    \caption{Interviewed participants}
    \label{tab:live_interview_ppl}
\end{table}

\begin{figure}[tb]
	\centering
	\includegraphics[width=\columnwidth]{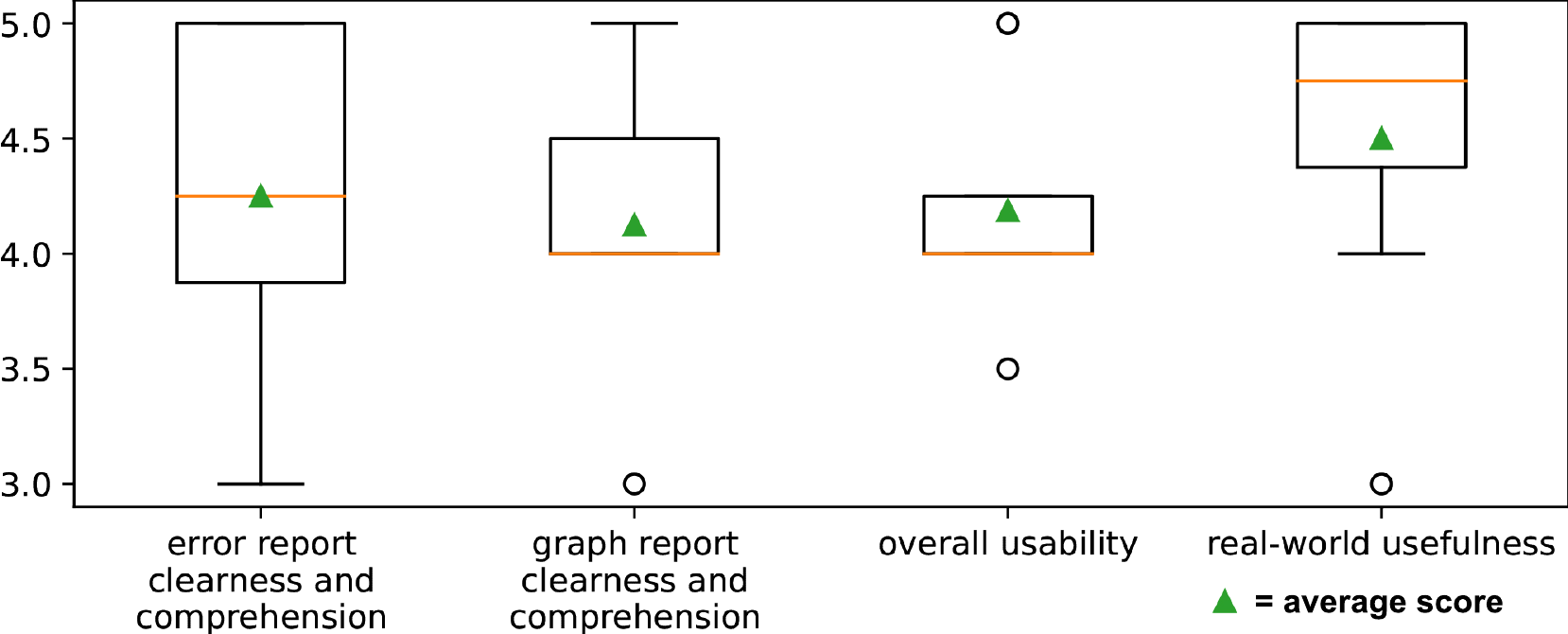}
	\caption{User evaluation result}
	\label{fig:Evaluation result} 
\end{figure}

After performing the interviews, we aggregated the scores given by the participants. The result is shown in Figure~\ref{fig:Evaluation result}. 
We found that the overall average score (represented by a green triangle) that the research participants evaluated the usability of the system in terms of \textit{clearness and comprehension of the error report} at 4.25 out of 5. 
The average score for the \textit{clearness and comprehension of the graph report} is 4.1. 
Furthermore, Microusity received an average score of 4.1 in the \textit{overall system usability}. Lastly, for the \textit{real-world usefulness}, the average score is 4.5. 


We also asked them what they liked and disliked about the tool. Five participants agreed that Microusity's error report, which indicates which service has an error, is useful. They believed the tool can assist them in determining which service causes API security issues. It saves them time compared to manually identifying the service that causes the problem. Moreover, the participants give positive comments such as categorization of HTTP error types, working with more than one service, showing API test coverage, being open-source, and using a fuzzing technique that can discover more bugs than traditional or manual testing.
However, three participants reported that the tool has no getting-start instruction and high learning curve, which caused them to struggle with learning how to start using the tool. The other issues include requiring a lot of settings and pre-requisite knowledge about RESTler, maintaining the tool, in the long run, can be problematic, lacks of filtering options in the error report, and not presenting a guide on how to solve the detected security issues. We plan to improve the tool based on their comments in our future work. 

The user evaluation reveals that the Microusity system is useful for software practitioners that work and maintain BFF microservice systems. 
The participants found the error and the graph reports useful and support the debugging of one-to-many requests created by BFF. 

\section{Related Work}
Chondamrongkul et al.~\cite{Chondamrongkul2020} proposed a security analysis approach for the microservice architecture model. The approach analyzes the microservice model defined using Ontology Web Language and Architecture Description Language and identifies security issues. Compare to their work, Microusity performs the API testing on the actual microservice system instead of the model. 
RESTTESTGEN~\cite{Viglianisi2020} and QuickREST~\cite{Karlsson2020} are automated testing tools for RESTful APIs. Similar to Microusity, the two tools rely on the specification of the APIs (e.g., Swagger or OpenAPI) to generate test inputs and locate the errors based on the HTTP response status code. EvoMaster~\cite{Arcuri2021} is a search-based white-box automated testing tool that can be applied to RESTful API testing. 

These existing tools can mostly test the RESTful APIs at the endpoints but do not trace the requests behind the endpoints. In contrast, Microusity deploys RESTler and Zeek and monitors the requests and responses. Therefore, Microusity can trace errors from the request and response sequence across the BFF as well as their back-end microservices. Microusity then provides easy-to-understand reports for the mapping of the service which causes such errors. 

\section{Conclusion}
We propose Microusity, an automated RESTful API  testing tool for BFF microservice systems using stateful fuzzing and main-request and sub-request mapping. The tool detects the requests that create HTTP 500--599 responses with exception messages from the back-end microservices and creates an error report and a graph report to aid the developers in comprehensing the issues. The evaluation with eight practitioners shows that the participants found the tool easy to use and useful during their development and maintenance of microservices.

\bibliographystyle{IEEEtran}
\bibliography{references}

\end{document}